\documentclass{aastex631}

\newcommand{\pgalf}{\texttt{PGalF}}
\newcommand{\CEAGLE}{\texttt{C-EAGLE}}
\definecolor{red}{RGB}{250,0,0}
\definecolor{revblue}{RGB}{0,60,220}

\usepackage{ulem}
\usepackage{multirow}

\shorttitle{ICL in an SIDM universe}
\shortauthors{Yoo et al.}

\graphicspath{{./}{figures/}}

\begin{document}

\title{Intracluster Light as a Probe for Dark Matter: Exploring Self-interacting Dark Matter and Cold Dark Matter with C-EAGLE Sims}
\correspondingauthor{Cristiano G.\ Sabiu}
\email{csabiu@uos.ac.kr}

\author[0000-0002-6841-8329]{Jaewon Yoo}
\affiliation{Korea Astronomy and Space Science Institute (KASI), Daedeokdae-ro, Daejeon 34055, Korea}
\affiliation{Quantum Universe Center, Korea Institute for Advanced Study, 85 Hoegi-ro, Dongdaemun-gu, Seoul 02455, Korea}

\author[0000-0002-7542-0355]{Ellen L.\ Sirks}
\affiliation{School of Physics, The University of Sydney and ARC Centre of Excellence for Dark Matter Particle Physics, Sydney, NSW, 2052, Australia}
\affiliation{Departamento de F\'isica Te\'orica, Facultad de Ciencias M-8, Universidad Aut\'onoma de Madrid, 28049 Madrid, Spain}

\author[0000-0002-5513-5303]{Cristiano G.\ Sabiu}
\affiliation{Natural Science Research Institute (NSRI), University of Seoul, Seoul 02504, Korea}

\author[0000-0001-9521-6397]{Changbom Park}
\affiliation{School of Physics, Korea Institute for Advanced Study, 85 Hoegiro, Dongdaemun-gu, Seoul 02455, Korea}

\author[0000-0002-6810-1778]{Jaehyun Lee}
\affiliation{Korea Astronomy and Space Science Institute (KASI), Daedeokdae-ro, Daejeon 34055, Korea}
\affiliation{School of Physics, Korea Institute for Advanced Study, 85 Hoegiro, Dongdaemun-gu, Seoul 02455, Korea}

\author[0000-0001-5427-4515]{Ankit Singh}
\affiliation{School of Physics and Astronomy, University of Nottingham, University Park, Nottingham NG7 2RD, United Kingdom}

\author[0000-0002-4391-2275]{Juhan Kim}
\affiliation{Center for Advanced Computation, Korea Institute for Advanced Study, 85 Hoegiro, Dongdaemun-gu, Seoul 02455, Korea}

\author[0000-0002-9434-5936]{Jongwan Ko}
\affiliation{Korea Astronomy and Space Science Institute (KASI), Daedeokdae-ro, Daejeon 34055, Korea}
\affiliation{University of Science and Technology (UST), Gajeong-ro, Daejeon 34113, Korea}

\begin{abstract}
In this pilot study, we investigate whether intracluster light (ICL) can serve as an observational discriminator of dark matter physics. The self-interacting dark matter (SIDM) model has gained increasing attention as a possible resolution to small-scale discrepancies between collisionless cold dark matter (CDM) simulations and observations, predicting distinct tidal interaction histories within galaxy clusters. We analyze Cluster-EAGLE zoom-in galaxy clusters re-simulated from identical initial conditions in both CDM and SIDM frameworks. The morphological similarity between dark matter and multiple baryonic tracers (gas, all stars, galaxies, and the combined brightest cluster galaxy plus ICL (BCG+ICL)) is quantified using the Weighted Overlap Coefficient, a contour-overlap statistic.
We find that dark matter is traced most accurately by BCG+ICL, followed by gas, all stars, and galaxies. %The BCG+ICL component remains a robust tracer even at high redshift, while gas initially traces dark matter poorly but improves over time, eventually approaching the performance of BCG+ICL. 
Although the differences are modest, they are systematic: the collisionless tracers (all stars, galaxies, and BCG+ICL) show better correspondence with the dark matter distribution in the CDM, while the relative advantage of BCG+ICL over gas is reduced in SIDM. This reduction is driven primarily by weaker BCG+ICL--dark matter correspondence in SIDM. Qualitatively, this behavior is consistent with the expectation that collisionless stellar components more naturally trace collisionless dark matter in CDM, whereas self-interactions in SIDM introduce an effective viscosity that weakens the dominance of BCG+ICL as a dark matter tracer relative to gas. 
Our results demonstrate the potential of ICL as a novel observational probe of dark matter physics and provide a first step toward using diffuse cluster light to constrain the nature of dark matter.
\end{abstract}


\section{Introduction} \label{sec:intro}
Galaxy clusters serve as powerful laboratories for investigating the nature of dark matter, as they are the most massive gravitationally bound systems in the universe, dominated by dark matter in both mass and dynamics. Their deep gravitational potentials shape the spatial distributions of baryonic components, providing a unique environment to probe the interplay between visible matter and the underlying dark matter framework, as well as the dynamics driving galaxy evolution \citep{2009ApJ...699.1595P}. In particular, the large-scale gravitational effects within clusters amplify signatures of dark matter, making them ideal environments to test cosmological models, constrain the properties of dark matter, and examine deviations from standard gravitational physics.

Rich galaxy clusters often host a massive brightest cluster galaxy (BCG), typically located near the bottom of the cluster's gravitational potential well and surrounded by intracluster light (ICL)—diffuse light from stars not gravitationally bound to any individual cluster galaxy \citep[e.g.,][]{2007AJ....134..466K, 2005MNRAS.358..949Z, 2017ApJ...834...16M, 2009A&A...507..621C, 2021Galax...9...60C, 2022NatAs...6..308M}.
The ICL is thought to originate from various dynamical interactions, including galaxy–galaxy encounters and tidal stripping by the cluster potential~\citep{2006ApJ...648..936R, 2014MNRAS.437.3787C, 2024AJ....167....7C, 2024ApJ...969..142C, 2025A&A...703A..85C, 2025MNRAS.538..622G}. As such, it encodes the cumulative effects of mergers and accretion events predicted by the hierarchical structure formation scenario in the $\Lambda$-cold dark matter (CDM) framework. Being collisionless and dynamically governed by the cluster’s global gravitational potential, the ICL is increasingly recognized as a promising visible tracer of dark matter \citep{2019MNRAS.482.2838M, 2020MNRAS.494.1859A, 2024A&A...683A..59C, 2022ApJS..261...28Y, 2024ApJ...965..145Y, 2025ApJ...988..229Y}. 

Recent studies have extended the comparison between the ICL and dark matter beyond spatial morphology, exploring their dynamical properties such as orbital energy \citep{2025MNRAS.539.2279B}. These works find that ICL stars tend to occupy more radial and lower-energy orbits than dark matter, resulting in a steeper BCG+ICL density profile—an effect also reported observationally \citep{2023A&A...679A.159D}. This suggests that while the ICL may act as a biased tracer of the underlying dark matter distribution, the difference in slope between the two components appears systematic. Indeed, \citet{2024ApJ...965..145Y} demonstrated a quantitative relation between their profiles, suggesting that the dark matter distribution can potentially be reconstructed from the BCG+ICL profile. Therefore, the ICL may serve as a biased but calibratable tracer of dark matter, even in the one-dimensional profile perspective.

The self-interacting dark matter (SIDM) model has gained increasing interest in recent years. From a theoretical standpoint, there is no fundamental reason to assume that dark matter particles are entirely collisionless \citep{2000PhRvL..84.3760S, 2000ApJ...534L.143B}, and weak self-interactions naturally arise in several particle physics frameworks proposed for dark matter \citep{2018PhR...730....1T}. On the observational side, SIDM offers a potential resolution to long-standing discrepancies between cold dark matter-only simulation predictions and observations of dwarf and low-mass galaxies \citep{2015MNRAS.453...29E, 2017ARA&A..55..343B}, including the well-known \textit{missing satellite} and \textit{core-cusp problems}. On cluster scales, merging systems such as the Bullet Cluster provide important upper limits on the self-interaction cross-section \citep[e.g.,][]{2004ApJ...606..819M, 2008ApJ...679.1173R, 2015Sci...347.1462H, 2018ApJ...869..104W, 2023ApJ...954...36W}.

In an SIDM universe, galaxies falling into a cluster can experience additional mass loss due to collisions between their dark matter particles and those of the host cluster. This process, known as dark matter \textit{evaporation}, complements conventional tidal stripping and accelerates the disruption of satellite galaxies. As a result, simulations of galaxy clusters in an SIDM framework (for example, with a cross-section of $\sigma/m = 1~\mathrm{cm}^2/\mathrm{g}$) exhibit fewer surviving satellites and a higher rate of disruption, ultimately yielding approximately 25\% less total mass at the present day compared to their CDM counterparts evolved from identical initial conditions \citep{2022MNRAS.511.5927S}.

An interaction cross-section of $\sigma/m = 1~\mathrm{cm}^2/\mathrm{g}$ corresponds to a regime of significant self-interactions. Assuming an NFW density profile for a galaxy cluster with total mass $10^{15}M_{\odot}$, virial radius of 1\,Mpc, and concentration parameter $c \sim 5$, the estimated mean free path in the central region (within $0.01$~Mpc) is approximately $24$\,kpc. This is much smaller than the cluster size, implying that self-interactions may substantially alter the dynamics and structure in dense cluster cores.
We note that the adopted SIDM cross-section, $\sigma/m = 1~\mathrm{cm}^2/\mathrm{g}$, lies near the upper end of the range currently allowed by cluster-scale observations \citep{2004ApJ...606..819M, 2008ApJ...679.1173R, 2018ApJ...869..104W}, and is therefore chosen here as a representative case to highlight the maximal observable impact of self-interactions in this pilot study.

Such differences in tidal interaction histories, manifesting as enhanced stripping from galactic outskirts and increased disruption, are expected to alter both the quantity and phase-space structure of the ICL. Given that the ICL originates from stars that have been stripped or tidally disrupted, it inherently encodes the imprint of the dark matter–baryonic matter interplay. As such, comparing the spatial distribution of the ICL with that of dark matter offers a promising pathway to distinguish between dark matter models, including SIDM and CDM.

In Section~\ref{sec:DataMethod}, we describe the simulated dataset, outline our methodology, and discuss its implementation. 
Section~\ref{sec:analysis} presents the mass assembly histories of the galaxy clusters and details our analysis procedures. We present and interpret our main results in Section~\ref{sec:result}, and conclude with a summary in Section~\ref{sec:conclusion}.

\section{Data and Method} \label{sec:DataMethod}
In this section, we outline the simulations, tracer definitions, and similarity metric used in our analysis. We employ two clusters from the Cluster-EAGLE (\CEAGLE) suite \citep{2017MNRAS.470.4186B} re-simulated under both CDM and SIDM models, with sufficient resolution to resolve ICL. We then identify halos, subhalos, galaxies, and unbound components using the PSB-based finder \pgalf, defining the ICL as unbound stars and the BCG+ICL as their union with the central galaxy. Finally, we describe the \textit{Weighted Overlap Coefficient} \citep[WOC;][]{2022ApJS..261...28Y}, a contour–overlap statistic that quantifies the spatial correspondence between baryonic tracers and dark matter, providing a normalized measure of similarity robust to masking and projection effects.

\subsection{Cluster-EAGLE Simulation}
The \CEAGLE\ suite comprises zoom-in cosmological hydrodynamic simulations of 30 galaxy clusters with total masses exceeding $10^{14}~M_{\odot}$ in a $\Lambda$CDM universe. For this study, we utilize two of these clusters that have been re-simulated from identical initial conditions under an SIDM model \citep[see][for details]{2018MNRAS.476L..20R}.
The \CEAGLE\ simulation achieves a mass resolution of $9.7 \times 10^6~M_{\odot}$ for dark matter particles, $1.8 \times 10^6~M_{\odot}$ for stellar particles, and spatial resolution of 0.7\,kpc, providing sufficient resolution to robustly study the ICL component (see related discussions in \citealt{2024ApJ...965..145Y, 2024MNRAS.535.2375M}).

In the SIDM simulations, an isotropic and velocity-independent dark matter self-interaction cross-section of $\sigma/m = 1~\mathrm{cm}^2/\mathrm{g}$ is adopted. At each simulation timestep $\Delta t$, dark matter particles interact via elastic scattering with neighbors within a fixed radius of $h_\mathrm{SI} = 2.66\,h^{-1}\mathrm{ckpc}$, with a scattering probability given by
\begin{equation}
P_\mathrm{scat} = \frac{(\sigma/m) m_\mathrm{DM} \nu \Delta t}{\frac{4\pi}{3} h_\mathrm{SI}^3},
\end{equation}
where $\nu$ is the relative velocity between particles and $m_\mathrm{DM}$ is the dark matter particle mass \citep[for more details see][]{2017MNRAS.467.4719R}.

Each of the two galaxy clusters used in this study (\texttt{CE12} and \texttt{CE05}, see Figure~\ref{fig:cluster}) has been simulated under both CDM and SIDM scenarios, with 30 snapshots covering cosmic time from $z \sim 14$ to $z = 0$.

\subsection{Structure Identification and Cluster Components}
\label{sec:structure_merger_trees}
We identify halos and galaxies from the entire snapshots of \CEAGLE\ using the Physically Self-Bound (PSB)-based galaxy finder~\citep[\pgalf, see Appendix~A of][for details]{2023ApJ...951..137K}. First, we find FoF halos from a unified data structure that contains all the matter components of dark matter, gas, stars, and massive black holes by utilizing the adaptive FoF algorithm of \pgalf. Inside each FoF halo we then search for self-bound stellar subhalos: local density peaks of the stellar component are detected on a TSC-smoothed stellar density field (TSC cell size $4\,h^{-1}\,$ckpc and Gaussian smoothing length $8\,h^{-1}\,$ckpc, with a peak threshold of $2\times10^{3}\,h^{2}\,M_{\odot}\,$ckpc$^{-3}$), peaks closer than $4\,h^{-1}\,$ckpc are merged, and candidate members are iteratively pruned through the combined total-energy and tidal-radius test for four iterations until convergence. A subhalo is retained as a galaxy if it contains at least $30$ self-bound stellar particles.

Within a cluster-scale FoF halo we then separate its stellar content into three mutually exclusive components. The \emph{BCG} is defined as the self-bound stellar subhalo whose density peak lies closest to the global gravitational potential minimum of the FoF halo; in every snapshot of \texttt{CE12} and \texttt{CE05} analyzed here this subhalo is also the most stellar-massive subhalo in the FoF halo. All other self-bound stellar subhalos within the same FoF halo are classified as \emph{satellite galaxies}. All stellar particles that belong to the FoF halo but are not bound to any subhalo are classified as the \emph{ICL}. By construction the three components are disjoint and their sum reproduces the total stellar mass of the FoF halo, so the combined BCG$+$ICL mass is invariant under this split.

\begin{figure}
\begin{center}
\includegraphics[width=1.0\columnwidth]{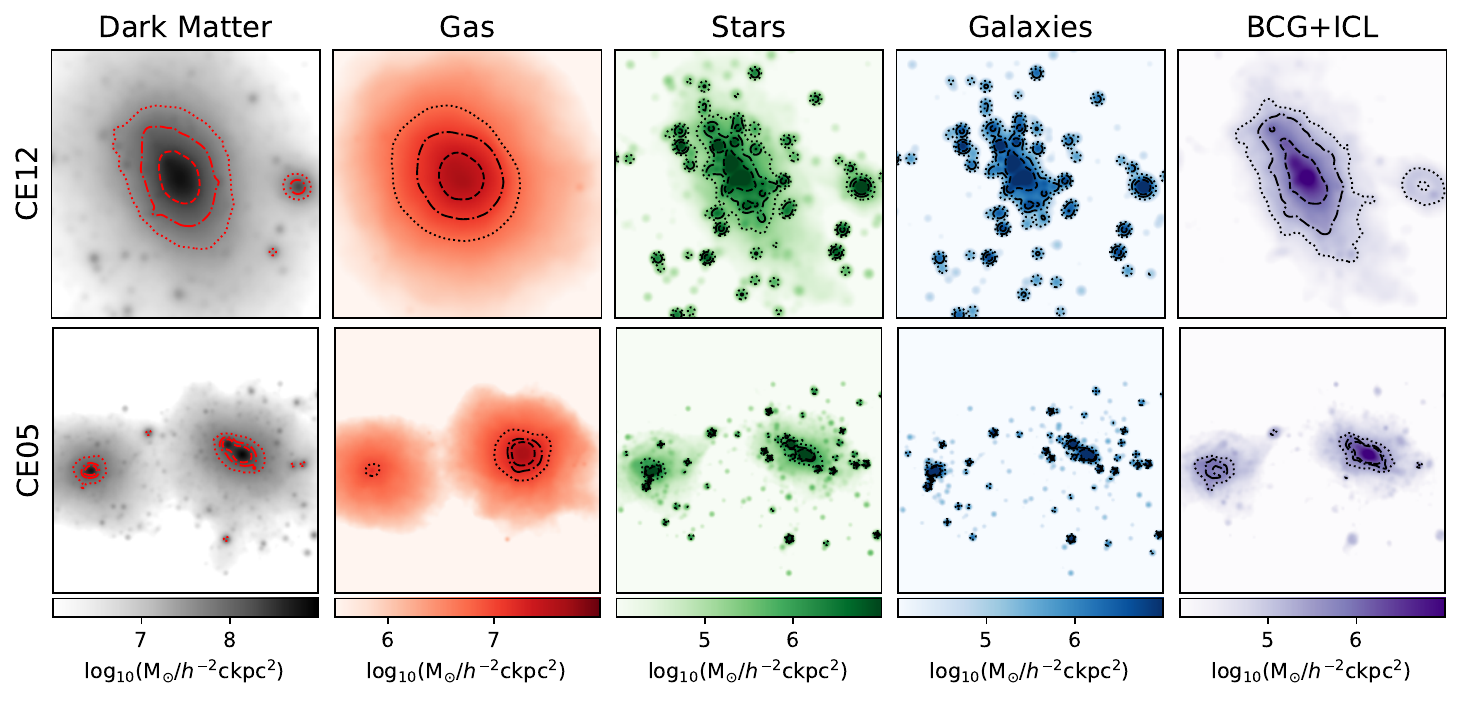}
\caption{Components of galaxy clusters in the \CEAGLE\ simulation at the final redshift ($z=0$) in CDM case. The upper row shows components of \texttt{CE12}, which is a relaxed galaxy cluster, and the lower row shows those of \texttt{CE05}, which is an unrelaxed galaxy cluster. 
The scales of images are 2 $\times$ 2 $h^{-2}$\,cMpc$^2$ for \texttt{CE12}, and 4 $\times$ 4 $h^{-2}$\,cMpc$^2$ for \texttt{CE05}. For each galaxy cluster, the dark matter, gas, all stars, galaxies, and BCG+ICL components are shown from the first column to the fifth column, respectively. Contours indicate the density levels of the azimuthally averaged radial density profiles at 0.1 (dashed), 0.2 (dash–dotted), and 0.3 (dotted) times the virial radius, at which the WOC, our similarity measure between components, is computed. Since no visually discernible morphological differences are observed between the CDM and SIDM realizations for either cluster, we present only the CDM case for clarity.
\label{fig:cluster}}
\end{center}
\end{figure}

During a merger between two halos, as long as the two progenitors are resolved as separate FoF halos each retains its own BCG under the above definition. Once they are linked into a single FoF halo \pgalf\ continues to resolve two self-bound stellar cores whenever the encounter is not yet fully coalesced: the core closer to the combined potential minimum is identified as the BCG of the merger remnant, while the other is classified as a massive satellite galaxy. Stellar particles that become tidally unbound during the encounter automatically migrate to the ICL at the first snapshot at which they fail the PSB test. When the two cores are no longer separable under the PSB criterion they are absorbed into a single BCG without any manual matching. The bottom panels of Figure~\ref{fig:cluster} illustrate this intermediate stage for \texttt{CE05}, in which the BCG, the most massive satellite, and the ICL are simultaneously present within the same FoF halo.

Because the combined BCG$+$ICL mass is invariant under the above split, the BCG$+$ICL mass fraction shown in the bottom row of Figure~\ref{fig:growth} and the WOC analysis in Sections~\ref{subsec:WOCmeasurement}--\ref{subsec:DMprobe}, which operate on BCG$+$ICL surface-density maps, are insensitive to the details of the BCG/ICL separation. The individual BCG and ICL mass curves (third and fourth rows of Figure~\ref{fig:growth}) and the corresponding fractions (fifth and sixth rows), however, do depend on the choice: because \pgalf\ classifies as ``BCG'' the entire self-bound central structure, which extends to $\sim300$--$500\,$kpc in these clusters, a fixed-aperture definition (for example $50\,$kpc around the BCG peak, widely used in observational BCG/ICL studies; \citealt{2015ApJ...806..207D,2018AstL...44....8K,2019ApJ...874..165Z}) would reassign the outer bound envelope to the ICL and shift $M_{\rm BCG}$ lower and $M_{\rm ICL}$ higher accordingly. We quantify this shift in Appendix~\ref{app:bcg_icl_robust}: the individual $M_{\rm BCG}$ curves move by up to $\sim2.7\,$dex between the PSB and the 50-kpc-aperture prescription, and the $M_{\rm ICL}$ curves by up to $\sim0.9\,$dex, while the combined $M_{\rm BCG}+M_{\rm ICL}$ and all CDM--SIDM offsets discussed in Sec.~\ref{sec:result} are preserved.

\subsection{Weighted Overlap Coefficient Method} \label{subsec:woc}
We quantify the similarity between two distributions of surface brightness or density by using the WOC, which ranges from 0 to 1. Measurement of the WOC of two distributions proceeds as follows.
First, the two maps are smoothed over the same angular or spatial scale, 
and the areas enclosed by the iso-density contours are measured at a set of threshold levels.
Then, in the comparison map, we find the threshold levels of contours that enclose the same areas as those in the reference map. The degree of overlap of these matched areas between two distributions is used to calculate the WOC with our choice of weights that take into account contour area and threshold density.

Specifically, suppose we are comparing a reference map $A$ with a comparison map $B$ using $n$ sets of matched regions $A_i$ and $B_i$ with ${\rm area}(A_i)={\rm area}(B_i)$. Let the $i=1$ level correspond to the highest level. 
The WOC is defined as
\begin{equation}
\mathrm{WOC}(A,B)=\frac{ \sum\limits_{i=1}^{n} f_{i} \left(w_{i}+w_{\rho_A,i}+w_{\rho_B,i}\right)}{\sum\limits_{i=1}^{n} 
\left(w_{i}+w_{\rho_A,i}+w_{\rho_B,i}\right)},
\label{eq:woc}
\end{equation}
where $f_{i}={\rm area}(A_i \bigcap B_i) /{\rm area}(A_i)$ is the fraction of overlapping area between the $i$-th contours of two maps, $w_i={\rm area}(A_i)^{-1}/\sum_j {\rm area}(A_j)^{-1}$ is a normalized weight weighing higher-level contour areas more,  and 
$w_{\rho_A,i}=\rho_{A,i}/\sum_j \rho_{A,j}$ and $w_{\rho_B,i}=\rho_{B,i}/\sum_j \rho_{B,j}$ are normalized weights giving more weights to higher threshold values.
$\rho_{A,i}$ and $\rho_{B,i}$ denote the density threshold at $i^{th}$ level on map $A$ and map $B$, respectively.

Therefore, contour areas and threshold levels are taken into account in our weighting system.
The WOC parameter quantifies the spatial correspondence of two maps rather than their relative signal strengths or the exact shape of their profiles. The WOC method works well even for disconnected regions and does not require computation of individual contours, making it less biased when working with masked maps with boundaries.
Detailed information on the methodology, tests on robustness, and comparisons with other methods can be found in \cite{2022ApJS..261...28Y}. 

\section{Analysis} \label{sec:analysis}
In this section, we (i) present the mass assembly histories of our two \CEAGLE\ clusters and (ii) describe how we construct the projected component maps and measure the WOC at each snapshot, comparing the spatial distribution similarity between the cluster components. We focus on \texttt{CE12} (relaxed) and \texttt{CE05} (disturbed), which are evolved from identical initial conditions under both CDM and SIDM prescriptions.

\subsection{Mass Growth History} \label{subsec:massgrowth}
\begin{figure}
\begin{center}
\includegraphics[width=0.89\columnwidth,trim={1.5cm 2.6cm 2.5cm 2.4cm},clip]{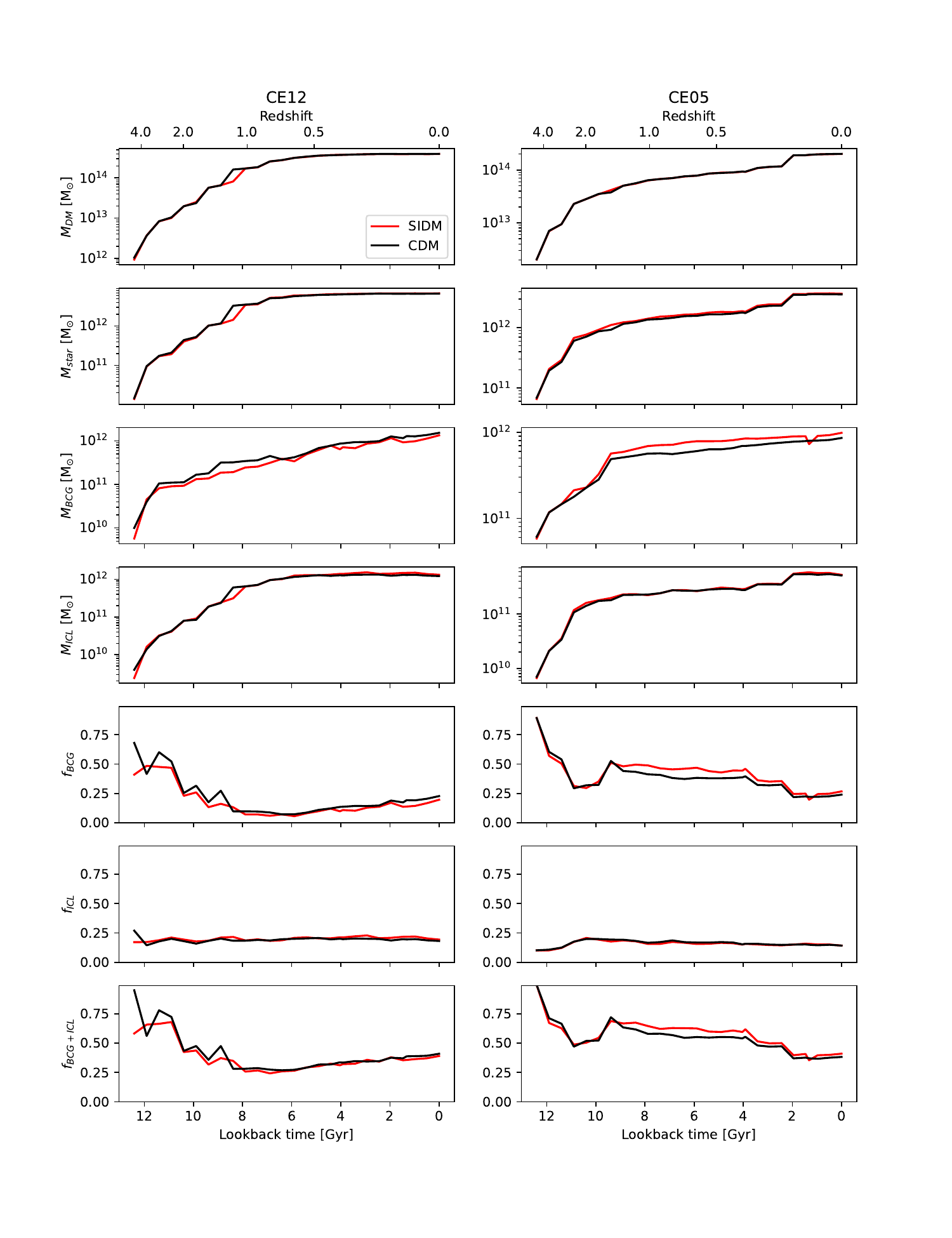}
\caption{Mass growth of various components in \texttt{CE12} (left) and \texttt{CE05} (right). From the first to the seventh row, we show the mass growth of dark matter, all stellar component, BCG, and ICL, followed by the stellar mass fraction evolution of the BCG, ICL, and BCG+ICL components. Results from the CDM universe are shown in black, and those from the SIDM universe in red. 
\label{fig:growth}}
\end{center}
\end{figure}

\begin{table}[t]
	\centering
 	\caption{ Cluster properties, half-mass epoch ($z_{m/2}$), last major merging redshift ($z_{\mathrm {LMM}}$)}
	\begin{tabular}{cc|ccccccc}
            \hline Cluster& Dark matter model&Total mass& Dark matter mass  & Stellar mass  &$\mathrm M_{200}$ &$\mathrm R_{200}$ & $z_{m/2}$ & $z_{\mathrm {LMM}}$ \\
            &&[${\mathrm M_\odot}$]&[${\mathrm M_\odot}$]&[${\mathrm M_\odot}$]&[${\mathrm M_\odot}$]&[$h^{-1}$cMpc]&&\\
            \hline \multirow{2}{*}{\texttt{CE12}} &CDM & $4.58\times 10^{14}$ & $3.89\times 10^{14}$  & $6.61\times 10^{12}$  & $3.96\times 10^{14}$  & 1.55&0.88&1.15  \\
             &SIDM & $4.57\times 10^{14}$&$3.89\times 10^{14}$ & $6.76\times 10^{12}$ & $3.91\times 10^{14}$ & 1.54&0.88&1.02\\
            \hline \multirow{2}{*}{\texttt{CE05}} &CDM & $2.33\times 10^{14}$&$2.00\times 10^{14}$  &  $3.55\times 10^{12}$ & $1.38\times 10^{14}$  & 1.09&0.32&0.16  \\
             &SIDM &  $2.32\times 10^{14}$&$2.00\times 10^{14}$ & $3.63\times 10^{12}$ & $1.36\times 10^{14}$ & 1.09&0.32&0.16\\
            \hline 
	\end{tabular}

	\label{tab:table2} 
\end{table}
Throughout this work, a galaxy cluster is consistently defined at each snapshot as the complete FoF-bound halo, including all gravitationally bound substructures, even when they appear morphologically disconnected. This definition is particularly important for merging systems, where the FoF halo can extend well beyond the virial radius (e.g., Figures~\ref{fig:cluster}, \ref{fig:woc_ce12_MM}, and \ref{fig:woc_ce5_MM}). Using this consistent cluster definition, we examined the mass growth histories of the two \CEAGLE\ clusters, \texttt{CE12} and \texttt{CE05}, for each structural component identified using our structure finder, \pgalf. We tracked the evolution of dark matter, all stellar particles, the BCG stellar component, and the ICL stellar component over a lookback time of 12–0 Gyr under both CDM and SIDM frameworks (Figure~\ref{fig:growth}). From these data, we derived the BCG, ICL, and BCG+ICL stellar mass fractions, defined as the ratio of each component’s stellar mass to the total stellar mass within the cluster, as functions of time. By the final snapshot at redshift zero, \texttt{CE12} and \texttt{CE05} have grown to total masses of approximately $4 \times 10^{14}~\mathrm M_{\odot}$ and $2 \times 10^{14}~\mathrm M_{\odot}$, respectively (Table~\ref{tab:table2}), showing no significant differences in overall mass growth between the two dark matter models.

We classify \texttt{CE12} as a relatively relaxed system and \texttt{CE05} as a relatively unrelaxed system primarily based on their half-mass epochs, which trace the global mass assembly history of the full cluster halo. The half-mass epoch is a widely used proxy for dynamical relaxation in simulation-based studies \citep{2022ApJS..261...28Y, 2024ApJ...965..145Y}. Compared with the 426 galaxy clusters at redshift zero in TNG300 \citep{2025ApJ...988..229Y}, \texttt{CE12} lies in the upper 16.5\% of the half-mass-epoch distribution, consistent with a more evolved system, whereas \texttt{CE05} lies in the lower 16.7\%, consistent with a dynamically younger system. This classification is also supported by their morphologies and recent mass growth histories: \texttt{CE12} shows a relatively round and symmetric morphology and an early phase of rapid growth followed by a long period of stability, while \texttt{CE05} displays a disturbed morphology associated with an ongoing major merger (Figure~\ref{fig:cluster}).

We note that this classification differs from that adopted by \citet{2018MNRAS.476L..20R,2022MNRAS.511.5927S}, who use the gas-based quantity $E_{\rm kin,500}/E_{\rm therm,500}$ measured within $R_{500}$. According to their criterion, $E_{\rm kin,500}/E_{\rm therm,500}<0.1$, both \texttt{CE12} and \texttt{CE05} would be classified as relaxed. However, that diagnostic probes the dynamical state of the hot gas in the central region, whereas our classification is intended to describe the global dynamical state of the entire cluster halo. Nevertheless, we emphasize that \texttt{CE12} and \texttt{CE05} are only two systems and therefore cannot represent the full diversity of relaxed and unrelaxed galaxy clusters.

The CDM and SIDM realizations show no visually discernible morphological differences for either cluster, so Figure~\ref{fig:cluster} presents only the CDM case for clarity. This lack of obvious visual differences motivates the use of the WOC, a quantitative metric capable of detecting subtle morphological variations that are difficult to identify by eye. In other words, it is precisely because the differences between CDM and SIDM are not readily apparent in visual inspection that a robust statistical measure such as WOC is required.

For both \texttt{CE12} and \texttt{CE05}, we find that whenever the dark matter mass exhibits a sharp increase, the total stellar mass and ICL mass also rise simultaneously. This temporal concurrence indicates that these episodes correspond to the accretion of galaxy groups into the main FoF halo. During such events, a substantial dark matter substructure—along with its constituent galaxies and pre-processed intra-group light—merges into the cluster, contributing simultaneously to both the dark matter and ICL components.

The BCG fraction is relatively high at early epochs ($z \sim 3$–2) and gradually decreases toward $z \sim 1$ in both clusters. At these redshifts, the total halo mass ranges from $10^{12}$ to $10^{13}~\mathrm{M_{\odot}}$, consistent with the scale of a galaxy group rather than a mature cluster. Subsequently, \texttt{CE12}, which evolves into a more relaxed system, exhibits a smooth and continuous increase in its BCG fraction from $z \sim 1$ (lookback time $\sim$ 8\,Gyr), reflecting steady accretion and gradual mass assembly thereafter.

The ICL fraction remains relatively constant at $\sim$15–20\% over the cosmological timescale, showing a nearly monotonic evolution. This trend is consistent with both observational findings \citep{2023Natur.613...37J} and our previous analysis using the Horizon Run 5 simulation \citep{2024ApJ...965..145Y}.

The BCG+ICL fraction is high at early epochs, primarily driven by the initially dominant BCG contribution. Since the ICL fraction remains relatively stable over most of the redshift range, the evolution of the BCG+ICL fraction is driven mainly by changes in the BCG fraction. Therefore, the BCG+ICL trend should be interpreted primarily as the evolution of the BCG-dominated central stellar component, rather than as a joint growth of both BCG and ICL. Nevertheless, we retain the BCG+ICL fraction as an important observationally motivated quantity, because separating the extended BCG envelope from the diffuse ICL is intrinsically ambiguous and method-dependent in real imaging data. For \texttt{CE12}, the BCG+ICL fraction decreases until $z \sim 1$, followed by a gradual increase toward lower redshifts, mainly following the evolution of the BCG component. In contrast, \texttt{CE05} shows a continuous decline in its BCG+ICL fraction over time. 
In particular, major merger events, identified by a sudden increase ($>33\%$) in total mass, coincide with sharp drops in the BCG+ICL fraction, reflecting the response of the central stellar component to merger-driven disturbances and rapid cluster mass growth.
%Previous studies have suggested that the BCG+ICL fraction reflects the dynamical state of the system \citep{2024ApJ...965..145Y, 2025ApJ...988..229Y}. The steady increase of this fraction from $z \sim 1$ to 0 in \texttt{CE12} indicates progressive relaxation and virialization. In \texttt{CE05}, by contrast, each major merger event of cluster -- identified by a sudden increase {\bf ($> 33\%$)} in total mass -- coincides with a sharp drop in the BCG+ICL fraction, implying that merger-driven disturbances disrupt the cluster’s relaxation state.

In the SIDM framework, we initially expected an enhanced production of ICL relative to the CDM case due to additional tidal stripping processes \citep{2022MNRAS.511.5927S}; however, no significant differences are observed in the total ICL or BCG+ICL fraction between the two models. This does not necessarily contradict previous findings of enhanced stellar stripping and higher satellite disruption rates in SIDM, because those differences are measured primarily for matched satellite galaxies. When considered at the level of the cluster as a whole, however, the difference is substantially reduced. The strongest SIDM effect is expected in the mass loss of satellite galaxies, particularly their dark-matter haloes and, to a lesser extent, their stellar components. Stars stripped from satellites are not removed from the total cluster stellar budget, but are instead redistributed into the diffuse stellar component and/or the extended envelope of the central galaxy. Since the pre-existing BCG+ICL reservoir is already massive, and because many of the additionally disrupted satellites in SIDM are relatively low in stellar mass, the extra stripped stellar mass may produce only a modest change in the integrated diffuse-light fraction. Our result therefore suggests that, while SIDM may alter the pathways by which stars are redistributed within clusters, its impact on the global ICL or BCG+ICL fraction can remain weak.

\subsection{WOC Measurement} \label{subsec:WOCmeasurement}
To assess how effectively different baryonic components trace the spatial distribution of dark matter in galaxy clusters, we analyze four distinct tracers: gas, all stellar particles, the galaxy stellar component, and the combined BCG+ICL stellar component. For each of the 30 simulation snapshots, we construct (i) dark matter density maps, (ii) gas surface density maps, (iii) stellar surface density maps including all stellar particles (BCG + satellite galaxies + ICL), (iv) galaxy surface density maps comprising the BCG and satellite galaxies, and (v) BCG+ICL surface density maps (see Figure~\ref{fig:cluster}). 

Each galaxy cluster component is projected onto a two-dimensional plane from 30 randomly sampled viewing directions on the sphere to mitigate projection-dependent biases. For each CDM–SIDM pair, the same lines of sight are used for both galaxy clusters and all components, ensuring that spatial distribution comparisons are performed in a consistent and unbiased manner. The resulting images are cropped to a scale determined by the overall extent of the structure, with an identical field of view applied to all components within each snapshot to ensure fair, consistent comparisons.

We smooth each surface density map with the same 1-$\sigma$ Gaussian smoothing kernel corresponding to 3 pixels. We then apply the WOC method to quantify the spatial correspondence between each baryonic component and the dark matter distribution, aiming to identify the most reliable dark matter tracer and to examine how this relationship evolves over cosmic time.

The WOC analysis was performed using radial bins corresponding to 0.1, 0.2, and 0.3 times the virial radius ($R_{200}$) of each cluster. In this calculation, the dark matter map serves as the reference against which the spatial distributions of gas, stars, galaxies, and the BCG+ICL components are compared. The center of each WOC calculation is aligned with the peak of the dark matter density, ensuring a physically consistent reference point. Because the contour levels of equal area are independently computed for each map, this approach does not require the assumption of a common center across all components, thereby allowing for a fair and robust comparison of their spatial distributions.

In this study, we calculated the WOC in 2D projected space to facilitate future comparisons with observed images. However, we also consider a 3D extension to the original WOC method, where overlapping area is now replaced by overlapping volume as a further test of the simulated data. We discuss and interpret both approaches in the Appendix~\ref{sec:3D}.

\section{Results and Discussions} \label{sec:result}
We now present the main results of our analysis, focusing on how different baryonic components trace the underlying dark matter distribution in both CDM and SIDM clusters. Using the WOC, we examine the evolution of tracer fidelity over cosmic time and assess how the relative performance of gas, stars, galaxies, and the BCG+ICL reflects both the dynamical state of the host cluster and the underlying properties of dark matter.

\subsection{Dark Matter Tracer} \label{subsec:DMtracer}
We begin by comparing, in a model-by-model and cluster-by-cluster manner, how closely each baryonic component follows the projected dark matter distribution. We use the time evolution of the WOC to identify the most faithful tracer and to assess how tracer performance changes as the clusters assemble and relax.
\begin{figure}
\begin{center}
    \includegraphics[width=1.0\columnwidth,trim={1.5cm 0.5cm 1.5cm 0.5cm},clip]{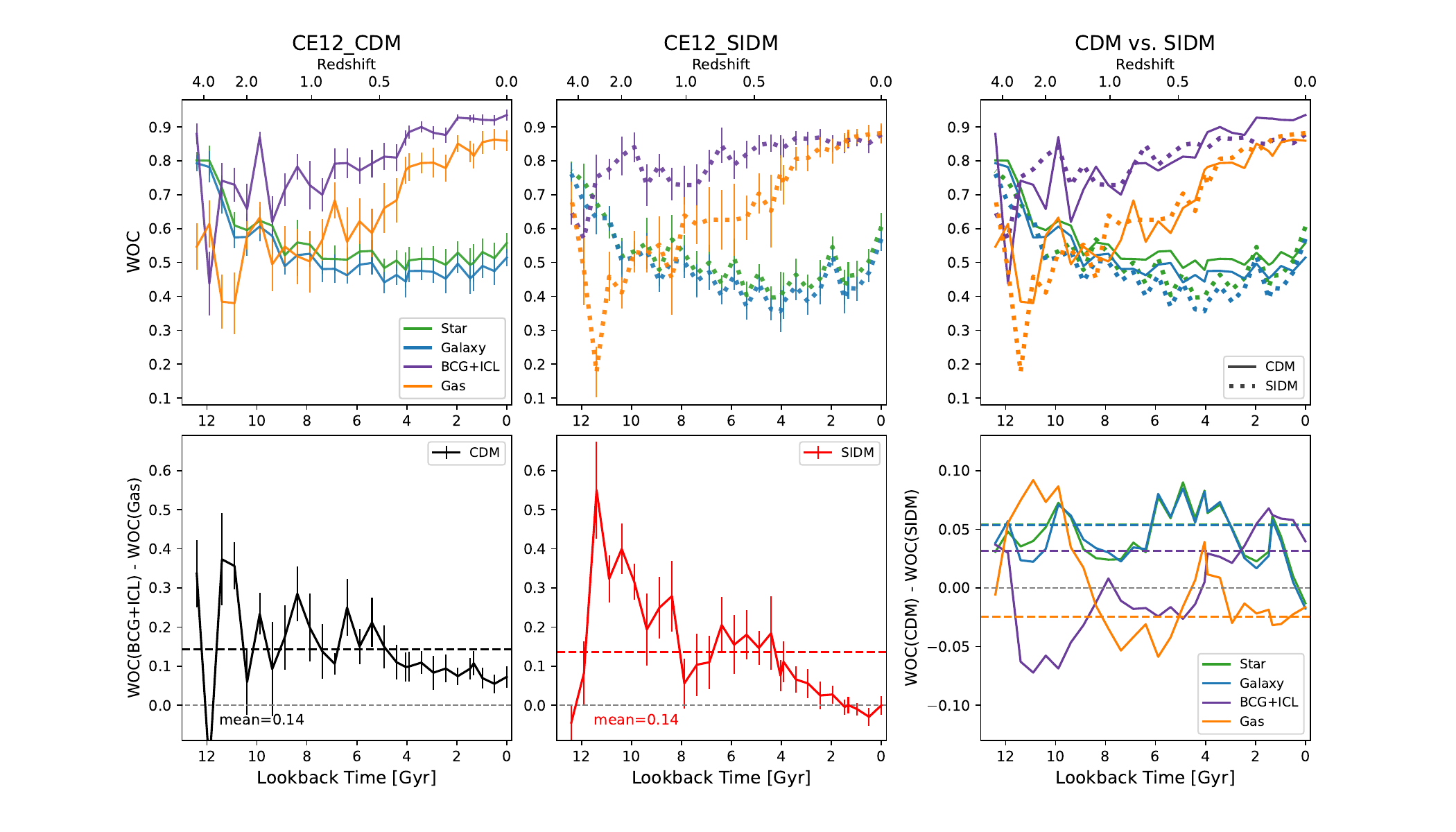}
\caption{
Evolution of the WOC for \texttt{CE12} in the CDM universe (left), SIDM universe (middle), and a comparison between CDM and SIDM (right).
(Top panels) Evolution of WOC values, representing the spatial distribution similarity of various components to the dark matter. Results for all stars, galaxies, BCG+ICL, and gas are shown in green, blue, purple, and orange, respectively. Solid lines indicate CDM results; dotted lines indicate SIDM.
(Bottom left and middle panels) Difference in WOC between BCG+ICL and gas, indicating the relative performance of these components in tracing dark matter, for CDM and SIDM, respectively. Colored horizontal dashed line indicates the time-averaged value. The gray dashed line denotes the zero reference level.
(Bottom right panel) WOC differences between the CDM and SIDM cases for each component, highlighting the impact of the dark matter model on spatial correlations. Horizontal dashed lines denote the median values for each component within $0 \leq z \leq 0.6$.
\label{fig:woc_ce12}}
\end{center}
\end{figure}

\begin{figure}
\begin{center}
    \includegraphics[width=1.0\columnwidth,trim={1.5cm 0.5cm 1.5cm 0.5cm},clip]{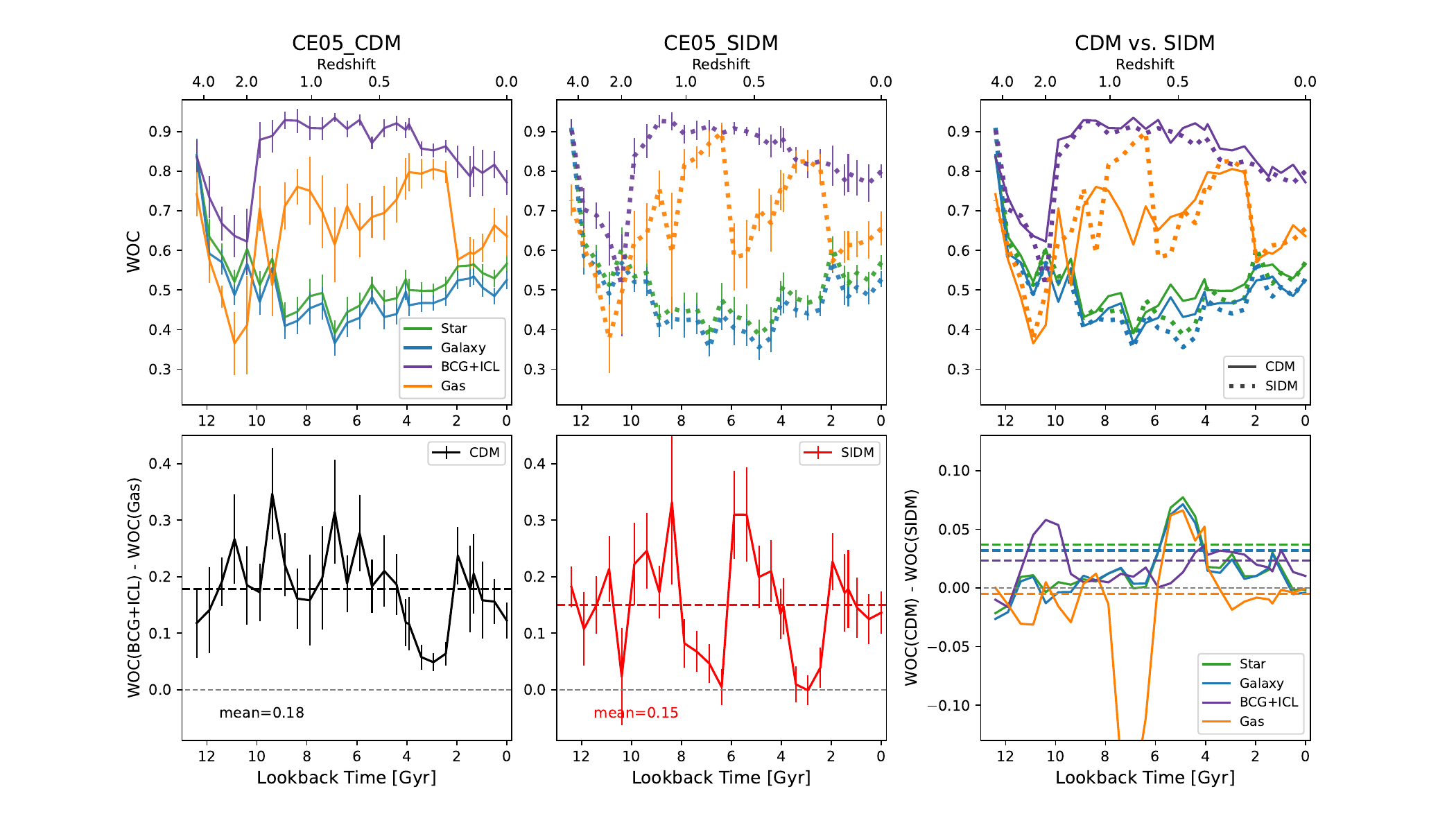}
\caption{Evolution of the WOC for \texttt{CE05}. The plotting conventions are identical to those in Figure~\ref{fig:woc_ce12}.
\label{fig:woc_ce5}}
\end{center}
\end{figure}
We show the WOC evolution for \texttt{CE12} (Figure \ref{fig:woc_ce12}) and \texttt{CE05}  (Figure \ref{fig:woc_ce5}), for each component relative to dark matter from lookback times of 12 to 0\,Gyr.
%In Figure \ref{fig:woc_ce12}, we show the WOC evolution for \texttt{CE12} in both DM scenarios for each component relative to dark matter from lookback times of 12 to 0\,Gyr. 
Higher WOC values indicate stronger spatial similarity and better correspondence with the dark matter distribution. The points and error bars represent the median and standard deviation, respectively, of the 30  measurements from the random projections. %The lower panel errors result from the error propagation of the two individual errors.
The values shown in the lower panels represent the median of the pair-wise differences, computed independently for each of the 30 random projections. For the left and middle panels, the differences are between the BCG+ICL and gas components, while for the right panel they are between the CDM and SIDM cases. Error bars indicate the corresponding standard deviations.

As we can see in the upper panels of both Figures \ref{fig:woc_ce12} and \ref{fig:woc_ce5}, except at high redshift, for both dark matter models (CDM and SIDM), the dark matter distribution is most accurately traced by the BCG+ICL component, followed by gas, all stars, and finally galaxies. Interestingly, for the relaxed galaxy cluster (\texttt{CE12}), the dark matter-tracing performance (i.e.\, the amplitude of WOC value) improves over time for the BCG+ICL and gas components, whereas it gradually declines for all stars and galaxies.

For \texttt{CE12}, the more dynamically relaxed cluster, the BCG+ICL consistently shows a strong correspondence with the dark matter distribution, even at high redshift. In contrast, the gas component correlates poorly with the dark matter distribution at early epochs, but its correlation increases, eventually reaching a level comparable to that of the BCG+ICL. This trend is consistent with results from the Horizon Run 5 simulation \citep{2024ApJ...965..145Y}. The decreasing residuals in the lower left panel in Figure~\ref{fig:woc_ce12}, which represent how BCG+ICL outperform compared to gas, further illustrate this convergence over time.
This tendency is less visible in \texttt{CE05}, the unrelaxed galaxy cluster case.

In a more detailed view, for example, around a lookback time of 11\,Gyr in \texttt{CE12} (upper panels in Figure~\ref{fig:woc_ce12}), the WOC value for the BCG+ICL remains high, or even increases, while that of the gas component drops sharply. This behavior reflects the physical processes occurring during a major merger event: the dissipative gas is shock-heated and scattered, making it less able to remain concentrated around the dark matter density peak, whereas the collisionless BCG+ICL component continues to follow the gravitational potential shaped by the dark matter distribution. 

Similarly, in \texttt{CE05} (upper panels in Figure~\ref{fig:woc_ce5}), after a lookback time of approximately 10\,Gyr, the WOC value for the BCG+ICL remains consistently high, whereas that of the gas component exhibits strong fluctuations. This behavior reflects the dynamically disturbed nature of \texttt{CE05}, where repeated merging and interaction events cause significant variations in the gas distribution, while the collisionless BCG+ICL continues to trace the underlying dark matter potential more stably.

\begin{figure}
\begin{center}
    \includegraphics[width=1.0\columnwidth,trim={0 0.3cm 0 0.4cm},clip]{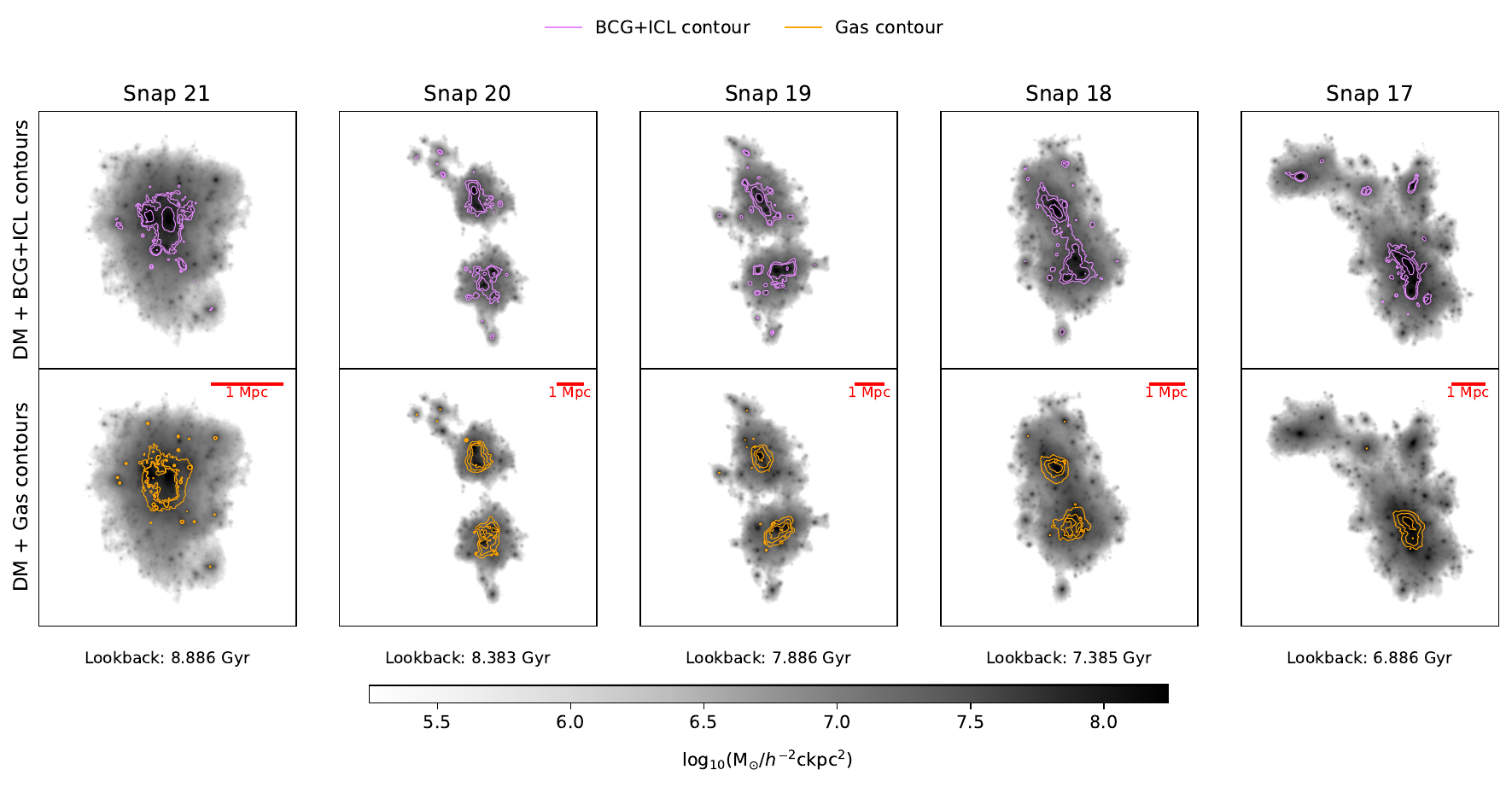}
\caption{ 
Snapshots of \texttt{CE12} around its major merger event (lookback time $\sim$ 8\,Gyr) in the CDM universe. Snapshots 21 through 17 are shown from left to right. (Top panels) The dark matter surface density maps are shown in the background, with BCG+ICL contours used in the WOC calculations overplotted in purple. (Bottom panels) The same dark matter maps are displayed, with gas contours used in the WOC calculations overplotted in orange. Because each snapshot adopts a different field of view to encompass the evolving structure, a 1\,Mpc scale bar is shown in each panel. The colorbar indicates the projected dark matter surface density.
\label{fig:woc_ce12_MM}}
\end{center}
\end{figure}

\begin{figure}
\begin{center}
    \includegraphics[width=1.0\columnwidth,trim={0 0.3cm 0 0.4cm},clip]{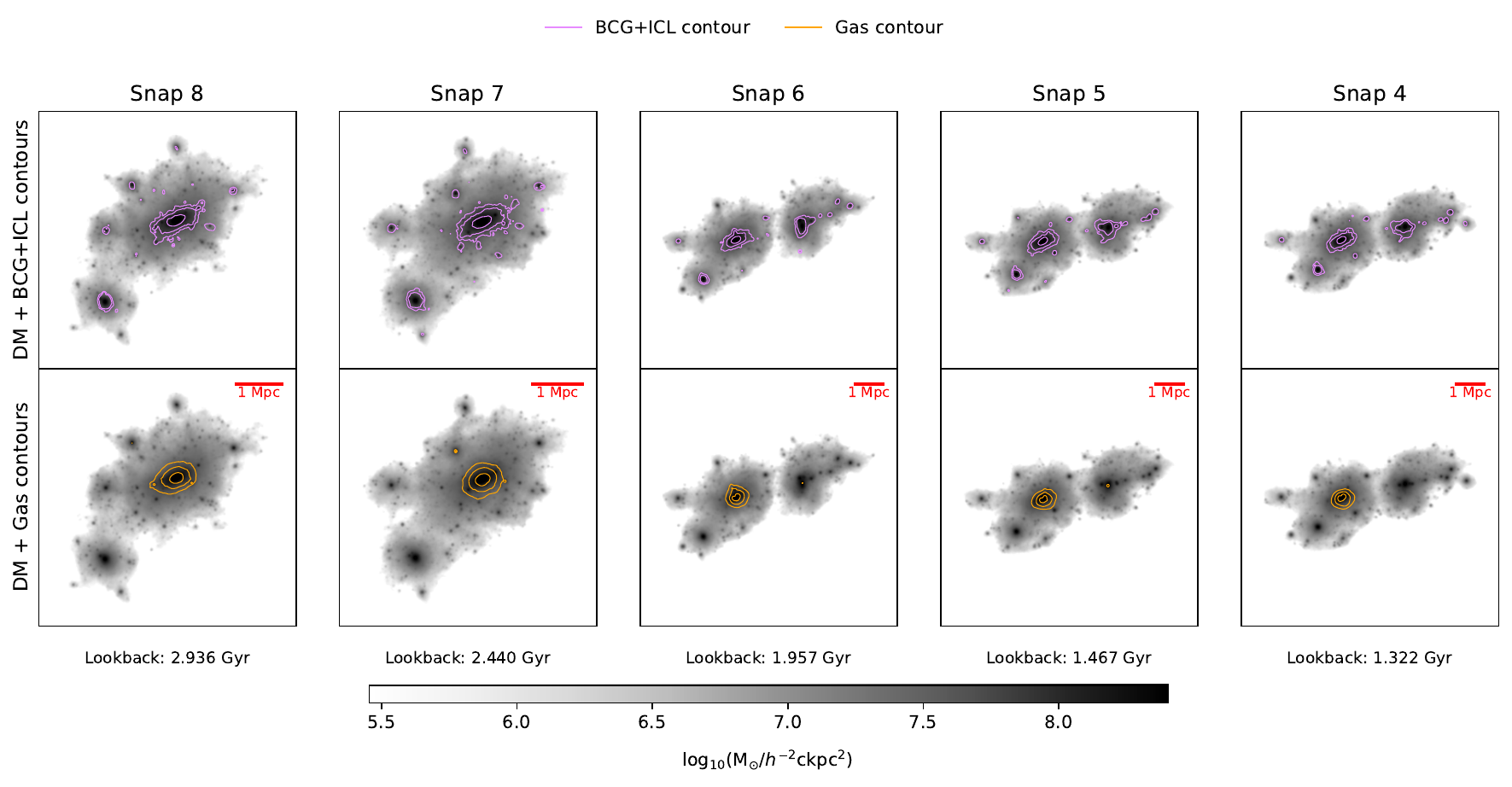}
\caption{Snapshots of \texttt{CE05} around its major merger event (lookback time $\sim$ 2\,Gyr) in the CDM universe. Snapshots 8 through 4 are shown from left to right. The plotting conventions are identical to those in Figure~\ref{fig:woc_ce12_MM}.
\label{fig:woc_ce5_MM}}
\end{center}
\end{figure}

Focusing on the major merger moments, at lookback times of $\sim$8 Gyr for \texttt{CE12} and $\sim$2 Gyr for \texttt{CE05} (see the top panels of Figure~\ref{fig:growth}), we visualize how effectively the BCG+ICL and gas components trace the dark matter distribution in Figures~\ref{fig:woc_ce12_MM} and~\ref{fig:woc_ce5_MM}. In Figure~\ref{fig:woc_ce12_MM}, we observe that at snapshot 18 the BCG+ICL rapidly follows the evolving dark matter morphology, while the gas remains separated into two distinct bodies, indicating a delayed merging response. Similarly, Figure~\ref{fig:woc_ce5_MM} shows that as a significant substructure (containing substantial dark matter) approaches the main cluster, the gas component is largely absent from the subhalo. This results in a relatively lower (and even sharply decreased for \texttt{CE05} at a lookback time of $\sim$2 Gyr as seen in Figure~\ref{fig:woc_ce5}) WOC value for gas compared to BCG+ICL, highlighting the distinct dynamical responses of collisional and collisionless components during mergers.

\citet{2022ApJS..261...28Y, 2024ApJ...965..145Y, 2025ApJ...988..229Y} demonstrated that the spatial distribution similarity between dark matter and the BCG+ICL component reflects the dynamical state of the cluster. Consistent with our previous studies, we find that the WOC(DM, BCG+ICL) of \texttt{CE12} increases after a lookback time of 8\,Gyr, as the cluster gradually relaxes over time. In contrast, for \texttt{CE05}, the WOC(DM, BCG+ICL) decreases after a lookback time of 4\,Gyr, coinciding with a recent major merger event that disrupts the system and increases its dynamical irregularity.

\subsection{Dark Matter Behavior in different Models} \label{subsec:DMbehavior}
In this subsection, we examine the differences between CDM and SIDM. 
Because of its self-interacting nature, SIDM effectively introduces a mild viscosity in the dark matter component. We quantify these differences by comparing how well BCG+ICL and gas trace the dark matter distribution, using the contrast between WOC(DM, BCG+ICL) and WOC(DM, gas) as our primary diagnostic.

Examining the residuals in the lower-left (CDM case; $\sigma/m =0~\mathrm{cm}^2/\mathrm{g}$) and lower-middle (SIDM case; $\sigma/m =1~\mathrm{cm}^2/\mathrm{g}$) panels of Figure~\ref{fig:woc_ce12}, where positive values indicate that BCG+ICL matches the dark matter distribution more closely than gas, we find that the residual declines more rapidly in the SIDM run than in the CDM run. However, this trend should not be interpreted as evidence that the gas distribution itself becomes a significantly better tracer of dark matter in SIDM. As shown in the upper-right panel, the WOC values for gas in the CDM and SIDM runs are broadly similar, especially over the last $\sim 4$ Gyr of lookback time. Instead, the reduced residual in the SIDM case appears to be driven mainly by the weaker relative performance of BCG+ICL as a dark matter tracer.

This behavior is qualitatively consistent with the expectation that, in CDM, the collisionless dark matter distribution is more closely followed by the collisionless stellar component, BCG+ICL. In SIDM, by contrast, self-interactions may alter the dark matter distribution in a way that reduces its spatial correspondence with BCG+ICL. Therefore, our result suggests that the relative advantage of BCG+ICL over gas as a dark matter tracer becomes weaker in the SIDM case, rather than demonstrating a clear enhancement of the gas--dark matter similarity.
In this framework, a larger interaction cross-section would enhance this effect, producing an even steeper slope. However, we would require more simulations with varying $\sigma/m$ to make any claim of this effect.  Nonetheless, we may speculate that the evolution rate of the gas–dark matter similarity (i.e., the slope of the residual) could be a function of cross-section, $f(\sigma)$, which may serve as an empirical indicator of the self-interaction cross-section. This effect, if found, may offer a potential means to constrain SIDM models through comparison with cosmological simulations and multi-wavelength observations. 
Especially, performing a statistical study of clusters over the redshift range $0<z<0.6$ and quantifying the slope of this evolutionary trend may offer an observational pathway to constrain the cross-section, the strength of dark-matter self-interactions.

In the bottom left and middle panels of Figures~\ref{fig:woc_ce12} and \ref{fig:woc_ce5}, the horizontal dashed lines indicate the time-averaged values of $\mathrm{WOC}(\mathrm{DM},\mathrm{BCG{+}ICL}) - \mathrm{WOC}(\mathrm{DM},\mathrm{gas})$ for the four cases considered: \texttt{CE12} in CDM, \texttt{CE12} in SIDM, \texttt{CE05} in CDM, and \texttt{CE05} in SIDM.
For the dynamically relaxed cluster \texttt{CE12}, the mean residuals are nearly identical in both dark matter models, with $\langle \Delta \mathrm{WOC} \rangle \simeq 0.15$, indicating that BCG+ICL and gas trace the dark matter distribution with comparable relative performance in both CDM and SIDM environments.
In contrast, for the dynamically disturbed cluster \texttt{CE05}, the mean residual in the CDM case ($\langle \Delta \mathrm{WOC} \rangle \simeq 0.19$) is noticeably larger than in the SIDM case ($\langle \Delta \mathrm{WOC} \rangle \simeq 0.15$). This suggests that during active merging, the collisionless BCG+ICL component outperforms the collisional gas in tracing the dark matter distribution more strongly in a CDM universe than in an SIDM one.
This behavior highlights that merging systems provide a particularly sensitive regime in which the distinction between collisionless and self-interacting dark matter can manifest through the relative tracer performance of BCG+ICL and gas.

The lower-right panel of Figure~\ref{fig:woc_ce12} and \ref{fig:woc_ce5} shows the difference in dark matter–tracing performance (WOC) between the two dark matter models (CDM – SIDM) for each component. In this panel, the residual is the difference between the solid and dotted curves in the upper panels. To reduce noise, the data were smoothed using a three-point boxcar average. The horizontal dashed lines indicate the median values of each component within the redshift range $0 < z < 0.6$.

Collisionless components (stars, galaxies, and BCG+ICL) exhibit higher WOC values in the CDM case compared to the SIDM case, i.e., WOC(CDM) $>$ WOC(SIDM), resulting in residuals above zero. In contrast, for the collisional gas component, WOC(CDM) $<$ WOC(SIDM)$,$ yielding residuals below zero, indicating that in the SIDM scenario, dark matter behaves more similarly to gas.
In the more relaxed system (\texttt{CE12}), the performance difference between CDM and SIDM is more pronounced than in the dynamically disturbed system (\texttt{CE05}). 

Among the cluster components, the sensitivity to the underlying dark matter model (quantified by the magnitude of deviation from zero in the lower right panels of Figures~\ref{fig:woc_ce12} and \ref{fig:woc_ce5}) follows the order: all stars $\sim$ galaxies $>$ BCG+ICL $>$ gas, for both \texttt{CE12} and \texttt{CE05}.
Thus, while both BCG+ICL and gas remain effective tracers of dark matter, the satellite galaxy containing components (all stars and galaxies) retain relatively stronger diagnostic value by exhibiting noticeable deviations between the CDM and SIDM cases, consistent with the findings of \citet{2022MNRAS.511.5927S}. This behavior may reflect the fact that dwarf and satellite galaxies are more sensitive to their surrounding environment, thereby encoding the influence of differing dark matter interaction properties more distinctly. However, observationally identifying dwarf and satellite galaxies with high spectroscopic completeness is challenging, especially in cluster environments. We therefore focus on BCG+ICL as a practical tracer: although its CDM--SIDM deviation is weaker than that of satellite-dominated components, it remains one of the cleanest tracers of dark matter morphology and is more observationally feasible for future low-surface-brightness imaging studies.

The observed differences (up to $\Delta \mathrm{WOC}\sim0.054$, indicated by the horizontal dashed lines in the lower-right panels of Figures~\ref{fig:woc_ce12} and \ref{fig:woc_ce5}) are similar to the scatter introduced by the 30 random projection directions ($\Delta \mathrm{WOC}\sim0.03-0.05$), and therefore should be interpreted with caution. Nevertheless, because each CDM--SIDM pair is re-simulated from the same initial conditions and analyzed along identical lines of sight, the pairwise comparison provides a controlled way to isolate differences arising from the dark matter model. Although the resulting CDM--SIDM differences are small and remain comparable to the line-of-sight scatter, the relative trends among cluster components still offer useful insight into how different tracers may respond to the underlying dark matter physics. %Nevertheless, because the CDM and SIDM comparisons for all components are performed along identical lines of sight, the relative differences among cluster components remain meaningful, preserving their comparative power for discriminating between dark matter models.

% \begin{table}[t]
% 	\centering
%  	\caption{ Pair-wise CDM-SIDM WOC differences over $0 \leq z \leq 0.6$.}
% 	\begin{tabular}{cc|cc}
%             \hline Cluster& Component & Median $\Delta \mathrm{WOC}$ & Std. \\
%             \hline \multirow{4}{*}{\texttt{CE12}} &Stars &  0.054 & 0.049 \\
%              &Galaxies &  0.054 & 0.051 \\
%              & BCG+ICL  &  0.031 & 0.035 \\
%              & Gas      & -0.025 & 0.036 \\
%             \hline \multirow{4}{*}{\texttt{CE05}} &Stars    &  0.037 & 0.036 \\
%             & Galaxies &  0.032 & 0.034 \\
%             & BCG+ICL  &  0.023 & 0.025 \\
%             & Gas      & -0.005 & 0.040 \\
%             \hline 
% 	\end{tabular}
% 	\label{tab:table3} 
% \end{table}

\subsection{ICL as Dark Matter Probe} \label{subsec:DMprobe}
We witnessed a difference in how various baryonic components trace the underlying dark matter in the CDM and SIDM universes. The relative tracing performance of gas, galaxies, and BCG+ICL encodes information about the dynamical coupling between baryons and dark matter, which varies depending on the presence or absence of dark matter self-interactions. Specifically, gas traces dark matter more closely in SIDM, and its performance improvement was faster in SIDM, while the BCG+ICL component provides a better match in CDM. This contrast reflects how self-interactions redistribute the dark matter density and velocity structure, modifying the baryonic response and thus the observable morphology of each component.

These systematic differences, if measurable through observations, could serve as an indirect probe of dark matter microphysics. For example, a stronger spatial alignment between dark matter and gas than between dark matter and ICL in relaxed galaxy clusters could indicate the presence of collisional dark matter, as expected in SIDM scenarios where self-interactions isotropize the dark matter distribution and make it behave more like a fluid. 
Indeed, we observed in the Horizon Run 5 simulation that the dark matter morphology lies between that of the gas and BCG+ICL components; more structured and detailed than the gas distribution, yet smoother and less clumpy than the BCG+ICL \citep{2024ApJ...965..145Y}. This intermediate nature of dark matter morphology highlights that neither the purely collisional (gas) nor the collisionless but mass-segregated (stellar) components alone can fully capture the spatial complexity of the dark matter distribution. Consequently, comparing both tracers simultaneously provides a more complete diagnostic of the underlying dark matter physics, particularly in distinguishing between collisionless CDM and self-interacting SIDM frameworks.
Quantitatively, morphological residuals such as $\mathrm{WOC(DM, gas) - WOC(DM, ICL)}$ could serve as diagnostic parameters capable of statistically distinguishing between CDM and SIDM clusters.

To translate this relative tracing efficiency into a quantitative constraint on the self-interaction cross-section, several steps are required. First, a large statistical sample of simulated clusters across a range of SIDM cross-sections (e.g., 0.1, 0.5, 1, $5~\mathrm{cm}^2/\mathrm{g}$) must be analyzed to calibrate how the tracer–dark-matter similarity varies with $\sigma/m$. Second, observational validation would demand high-quality, multiwavelength datasets: deep optical imaging for ICL, X-ray or SZ maps for gas, and gravitational lensing maps for dark matter. We may also need to consider that gravitational lensing is sensitive to the total matter component, not dark matter directly (which makes up $\sim$ 85\% of the cluster mass). Likewise, X-ray observations trace only the hot gas component, while most of the gas is indeed in this hot phase, a small fraction of cold gas is confined to galaxies. Finally, a consistent definition of spatial-similarity metrics, such as the WOC, is essential for comparing simulations and observations on equal footing.

In summary, the relative difference in dark matter–tracing performance between gas and BCG+ICL under CDM and SIDM frameworks provides a promising new diagnostic of dark matter physics. If systematically calibrated and applied to observational data, this difference could offer an indirect yet quantitative constraint on the strength of dark matter self-interactions, opening a new pathway to probe dark matter models through the joint analysis of ICL, X-ray, and lensing observations.

Future wide-field facilities such as Rubin’s Legacy Survey of Space and Time (LSST) will deliver statistically robust samples of galaxy clusters \citep{2009arXiv0912.0201L, 2020arXiv200111067B}, with deep optical imaging complemented by weak- and strong lensing mass maps that trace the 
total matter distribution, which in galaxy clusters is dominated by dark matter. When combined with X-ray observations, which trace the hot gas component and are particularly well suited to clusters where most of the gas resides in this hot phase, these data will provide an unprecedented multiwavelength view of cluster structure and evolution, enabling stringent tests of dark matter physics. Comparing observations with CDM and SIDM simulations suggests that the WOC may serve as a useful diagnostic of the underlying dark matter model.
If the WOC of gas proves to be a more reliable tracer of dark matter in relaxed clusters at $z \sim 0$, it would favor an SIDM-like universe. However, if the BCG+ICL component is found to trace dark matter more reliably, this would favor a conventional CDM-like scenario or an SIDM model with a smaller interaction cross-section than that adopted in the existing SIDM simulation (for example, regarding this study, $\sigma/m < 1~\mathrm{cm}^2/\mathrm{g}$). In such a case, the resulting behavior would be nearly indistinguishable from that of a purely collisionless CDM universe. However, our analysis is based on only two simulated clusters, and the observed differences are modest. A substantially larger sample spanning a range of cluster masses, dynamical states, and SIDM cross-sections will be required to determine whether this trend is statistically robust and can provide meaningful constraints on the nature of dark matter.

\section{Conclusions} \label{sec:conclusion}

In this work, we have explored how different baryonic components trace the underlying dark matter distribution in galaxy clusters, using two systems from the \CEAGLE\ simulations evolved under both CDM and SIDM prescriptions. By applying the Weighted Overlap Coefficient (WOC), we quantified the correspondence between dark matter and gas, stars, galaxies, and the combined BCG+ICL component across cosmic time. Our analysis shows that BCG+ICL consistently provides the most faithful tracer of dark matter in both models, with its fidelity improving toward lower redshifts, while galaxies and the total stellar population become progressively poorer tracers. Gas initially traces the dark matter poorly, but in the relaxed cluster case, its correspondence improves substantially over time and eventually approaches that of the BCG+ICL component. In the SIDM case, this convergence occurs more rapidly than in CDM, and gas modestly exceeds BCG+ICL by $z=0$. This behavior reflects the underlying physics: in CDM, collisionless dark matter is more naturally traced by the collisionless BCG+ICL, whereas in SIDM, self-interactions introduce an effective collisionality, that reduces the relative advantage of BCG+ICL over gas as a dark matter tracer.

The evolutionary patterns of BCG, ICL, and BCG+ICL fractions further highlight the sensitivity of diffuse stellar components to the dynamical state of their host clusters. The relaxed system, \texttt{CE12}, shows steady growth and an increasing BCG+ICL fraction after $z \sim 1$, while the disturbed system, \texttt{CE05}, displayed a disrupted growth history and declining fractions during merger episodes. These results reinforce the notion that diffuse light is not only a tracer of the cluster potential but also a marker of the system’s dynamical maturity.  

Although the two SIDM clusters analyzed here do not exhibit large statistical deviations from their CDM counterparts, the relative behavior of gas and BCG+ICL with respect to dark matter suggests a possible observational pathway for distinguishing dark matter models. Specifically, the slope of the evolutionary trend comparing WOC(DM, BCG+ICL) and  WOC(DM, gas) may become steeper for larger dark matter self-interaction cross-sections. %{\bf However, stronger evidence for this trend will require future SIDM simulations with larger cluster samples and a broader range of cross-sections.}
We further find that dwarf and satellite galaxy components exhibit enhanced sensitivity to the underlying dark matter model, retaining diagnostic power despite their poorer overall tracing performance.

Future wide-field surveys such as Rubin’s LSST and KASI Deep Rolling Imaging Fast Telescope (K-DRIFT; \citealt{2026JKAS...59...67K, 2026JKAS...59...81Y}), together with strong and weak lensing reconstructions and complementary X-ray observations, will provide the multiwavelength data necessary to test these predictions. If gas consistently emerges as the better tracer of the mass distribution in relaxed clusters, this may point toward SIDM-like behavior, whereas the continued dominance of BCG+ICL as a tracer would be more compatible with CDM or SIDM models with smaller cross-sections.  

This study represents an exploratory step toward calibrating ICL as a probe of the dark sector. Expanding the analysis to larger cluster samples, broader ranges of mass and redshift, different dynamical states, and multiple self-interaction cross-sections will be crucial for assessing the robustness of these signatures. We plan to extend this study to galaxy clusters from The300 project, which will provide a larger cluster sample, broader mass coverage, and simulations with different self-interaction cross-sections. Exploring velocity-dependent cross-sections will also be important for connecting cluster-scale ICL observables to dark matter physics more generally.

A key next step is to disentangle the relative contributions of baryonic physics and the underlying dark matter model to the evolution of WOC. In particular, the physical drivers of rapid WOC changes should be examined by tracing cluster evolution across snapshots and comparing them with the properties of the BCG and central environment, such as supernova and AGN feedback activity, merger phase, recent accretion history, and the growth of the diffuse stellar component. Such analyses would clarify whether variations in the spatial correspondence between dark matter and baryonic tracers are primarily driven by dark matter self-interactions or by baryonic processes that reshape the central cluster potential.

Ultimately, the diffuse stellar component, together with its relative behavior compared to gas, may provide a promising and complementary way to probe dark matter physics in galaxy clusters.
 Further improvements in our understanding of the ICL and its relation to dark matter will be aided by forthcoming observational facilities and instruments. As observations continue to improve, testing and validating these results in real-world observations of galaxy clusters will be crucial.

\begin{acknowledgments}
 We would like to thank the anonymous referee whose careful and insightful comments allowed us to improve the presentation of this work significantly. The authors thank Yannick Bah\'e and David Barnes for making the CDM C-EAGLE data available, and Andrew Robertson for allowing us to use the SIDM C-EAGLE data. 
J.Y. acknowledges support from the Korea Astronomy and Space Science Institute under the R\&D program (Project No. 2026-1-831-03) supervised by the Korea AeroSpace Administration.
J.Y. was supported by a KIAS Individual Grant (QP089902) via the Quantum Universe Center at Korea Institute for Advanced Study.
C.G.S. acknowledges support from the Basic Science Research Program (2018R1A6A1A06024977) through Korea's NRF funded by the Ministry of Education.
E.L.S. was supported by the Australian Government through the Australian Research Council Centre of Excellence for Dark Matter Particle Physics (CDM, CE200100008), and acknowledges support from a Juan de la Cierva fellowship (JDC2024-054790-I) of the Spanish Ministery of Science, Innovation, and Universities (MICINN). 
C.P. is supported by KIAS Individual Grants (PG016903) at Korea Institute for Advanced Study and by the National Research Foundation of Korea (NRF) grant funded by the Korean government (MSIT; RS-2024-00360385).
J.L. is supported by the National Research Foundation of Korea (NRF-2021R1C1C2011626). 
This work is supported by the Center for Advanced Computation at Korea Institute for Advanced Study. 
J.Y. thanks Jeong-Gyu Kim for the useful discussion.
\end{acknowledgments}

\vspace{5mm}
\facilities{Cluster-EAGLE}

\software{astropy \citep{2013A&A...558A..33A,2018AJ....156..123A}
          }

\bibliography{SIDM}{}
\bibliographystyle{aasjournal}

\appendix

\section{Robustness of the BCG/ICL Separation}
\label{app:bcg_icl_robust}

In Section~\ref{sec:structure_merger_trees} we defined the BCG as the self-bound stellar subhalo returned by \pgalf\ (in every snapshot of \texttt{CE12} and \texttt{CE05} this coincides with the most stellar-massive subhalo) and the ICL as the remaining unbound stellar component of the host FoF halo. Because \pgalf\ determines subhalo membership by the combined total-energy and tidal-radius test rather than by a fixed radial cut, the BCG subhalo typically extends to $r\sim300$--$500$~physical kpc in these clusters and encloses both the compact central galaxy and its bound diffuse envelope. An aperture-based definition, by contrast, cuts the BCG at a fixed physical radius and reassigns the outer bound envelope to the ICL. Here we quantify how this choice propagates into the mass-growth histories shown in Figure~\ref{fig:growth}.

As a representative aperture we adopt $R_{\rm ap}=50\,$kpc (physical), which has been used in a range of recent observational and theoretical BCG studies \citep{2015ApJ...806..207D,2018AstL...44....8K,2019ApJ...874..165Z} and is broadly consistent with the BCG half-light region in the mass range spanned by \texttt{CE12} and \texttt{CE05}. We recompute $M_{\rm BCG}(t)$ and $M_{\rm ICL}(t)$ for both clusters in both the CDM and SIDM realisations using
\begin{equation}
M_{\rm BCG}^{\rm ap} = \sum_{\,i\in \mathrm{BCG\,subhalo}} \!\!m_{\star,i}\,\Theta(R_{\rm ap}-r_i) \;+\; \sum_{\,j\in \mathrm{unbound}} \!\!m_{\star,j}\,\Theta(R_{\rm ap}-r_j),
\label{eq:mbcg_ap}
\end{equation}
with $r_i,r_j$ the distances from the BCG density peak. Satellite subhalos never intrude into the aperture in our clusters, so their bound masses are unaffected. We define $M_{\rm ICL}^{\rm ap} \equiv (M_{\rm BCG}^{\rm PSB} + M_{\rm ICL}^{\rm PSB}) - M_{\rm BCG}^{\rm ap}$, so that $M_{\rm BCG}+M_{\rm ICL}$ is identical under the two prescriptions by construction.

\begin{figure}
\begin{center}
\includegraphics[width=1.0\columnwidth]{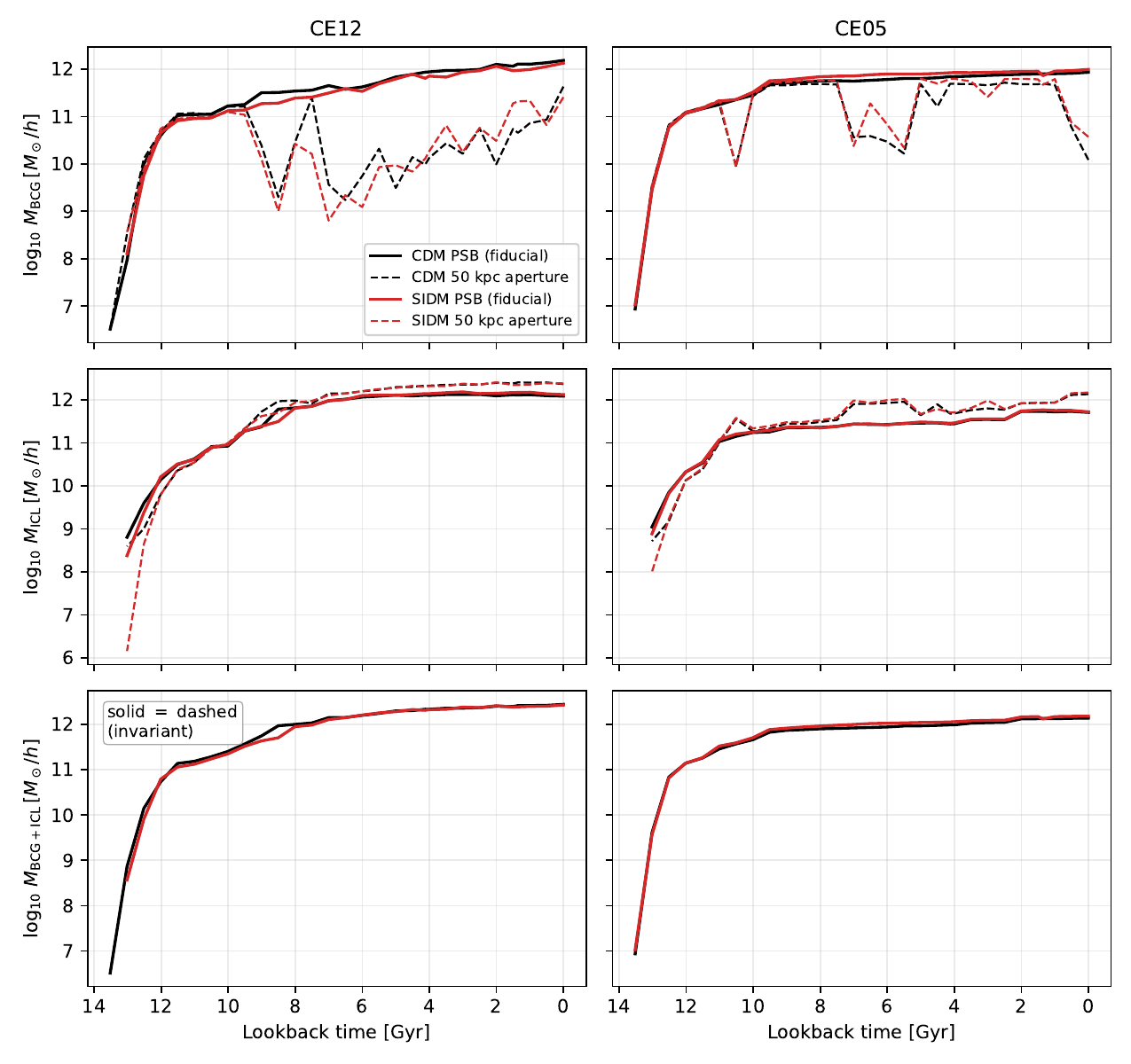}
\caption{ Robustness of the BCG/ICL separation. Left column: \texttt{CE12}; right column: \texttt{CE05}. Top row: $\log_{10} M_{\rm BCG}$; middle row: $\log_{10} M_{\rm ICL}$; bottom row: $\log_{10}(M_{\rm BCG}+M_{\rm ICL})$. Solid: fiducial \pgalf\ PSB definition used throughout this paper. Dashed: 50-kpc physical aperture around the BCG centre, which matches the convention adopted in recent BCG/ICL work \citep{2015ApJ...806..207D,2018AstL...44....8K,2019ApJ...874..165Z}. Black curves: CDM; red curves: SIDM. The individual $M_{\rm BCG}$ and $M_{\rm ICL}$ histories depend strongly on the definition, whereas the combined $M_{\rm BCG}+M_{\rm ICL}$ (bottom row) is invariant by construction and the two definitions yield indistinguishable curves. The WOC analysis in Sections~\ref{subsec:WOCmeasurement}--\ref{subsec:DMprobe} uses BCG$+$ICL maps and therefore inherits this invariance.
\label{fig:bcgicl_robust}}
\end{center}
\end{figure}

Figure~\ref{fig:bcgicl_robust} shows the result. The individual $M_{\rm BCG}$ curves differ substantially between the two prescriptions: because the \pgalf\ PSB subhalo captures a gravitationally bound envelope that extends well beyond the 50-kpc aperture, the aperture definition yields an $M_{\rm BCG}$ that is typically $0.2$--$1.3\,$dex (median) and up to $\sim2.7\,$dex lower than the PSB value, with the largest deviations occurring at $z<1$ when the BCG is most extended. Correspondingly, $M_{\rm ICL}^{\rm ap}$ is $0.2$--$0.3\,$dex (median) higher than $M_{\rm ICL}^{\rm PSB}$, since the bound envelope that the aperture excludes from the BCG is absorbed into the ICL. The combined $M_{\rm BCG}+M_{\rm ICL}$ curves (bottom row) are identical under both prescriptions to numerical precision.

This is important for the interpretation of the paper. The WOC similarity measurements of Sections~\ref{subsec:WOCmeasurement}--\ref{subsec:DMprobe} are computed on BCG$+$ICL surface-density maps and therefore depend only on the combined stellar distribution, which is invariant under the BCG/ICL split. The qualitative features visible in the individual $M_{\rm BCG}$ and $M_{\rm ICL}$ curves of Figure~\ref{fig:growth} --- notably the post-merger jumps discussed in Section~\ref{subsec:massgrowth} --- remain present under the aperture definition as well, but with a different overall normalisation: what the PSB definition labels as a rapid BCG growth event is labelled by the aperture definition as a corresponding ICL growth event, because stars tidally relocated to the bound envelope during the merger are beyond the 50-kpc radius. The CDM--SIDM offsets discussed in Section~\ref{sec:result} therefore persist under both prescriptions. We conclude that the main conclusions of this paper, which concern BCG$+$ICL morphology rather than an individual BCG--ICL decomposition, are robust against the choice of BCG/ICL separation criterion.

\section{2D vs 3D WOC}
\label{sec:3D}
\begin{figure}
\begin{center}
    \includegraphics[width=1.0\columnwidth,trim={0.3cm 0.3cm 0.3cm 1.5cm},clip]{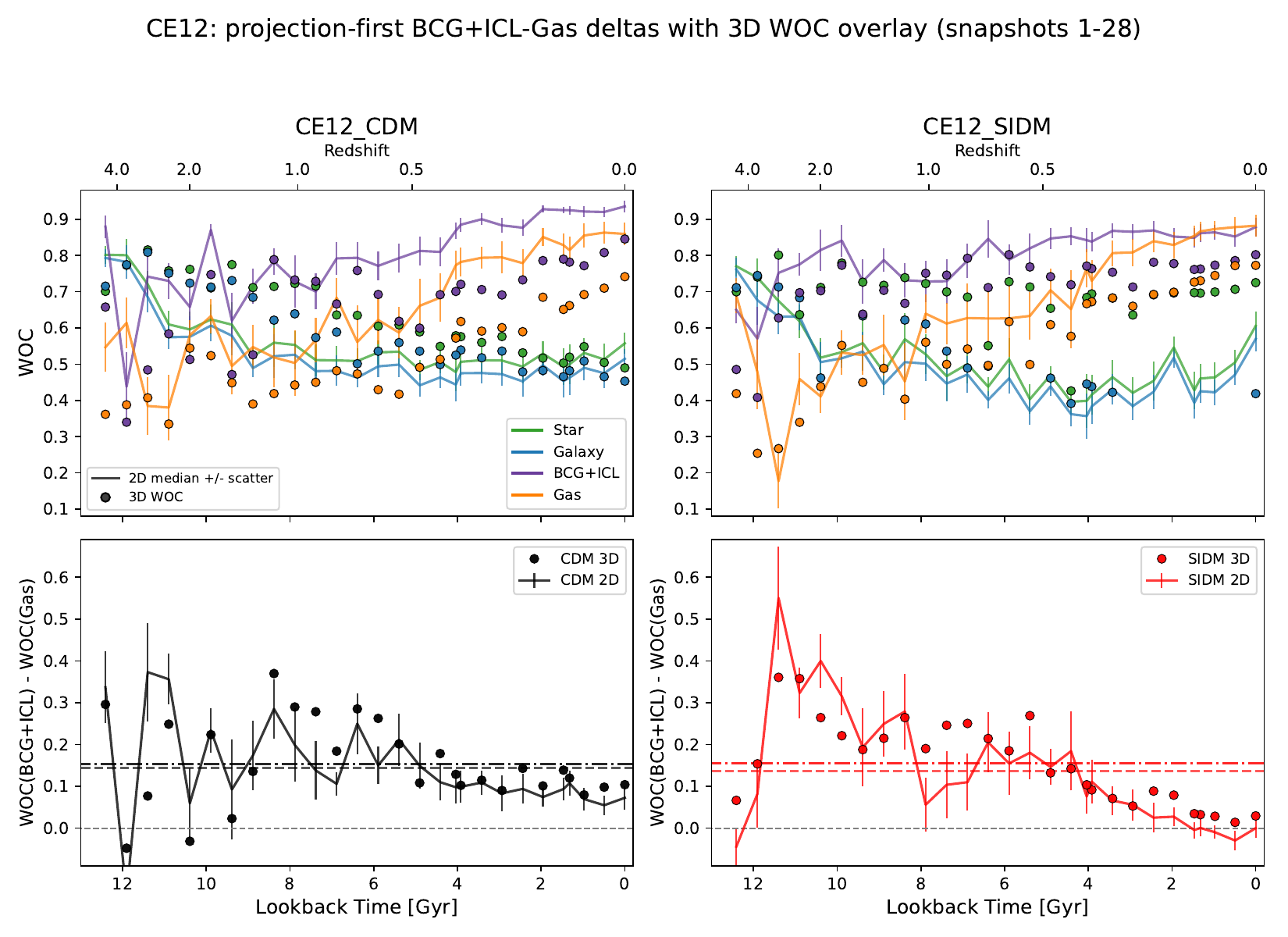}
\caption{
 Similar to Figure \ref{fig:woc_ce12} -- showing the 3D WOC results alongside the 2D WOC for cluster \texttt{CE12} in the CDM universe (left) and SIDM universe (right).
(Top panels) Evolution of 2D and 3D WOC values of various components to the dark matter. Results for all stars, galaxies, BCG+ICL, and gas are shown in green, blue, purple, and orange, respectively. Solid lines indicate 2D results; filled circles denote 3D WOC results.
(Bottom panels) Difference in WOC between BCG+ICL and gas for the 2D (lines) and 3D analysis (black cirles). Colored horizontal dashed and dash-dotted lines indicate the time-averaged values of the 2D and 3D results, respectively. The gray dashed line denotes the zero reference level.
\label{fig:3Dwoc_ce12}
}

\end{center}
\end{figure}

\begin{figure}
\begin{center}
    \includegraphics[width=1.0\columnwidth,trim={0.3cm 0.3cm 0.3cm 1.5cm},clip]{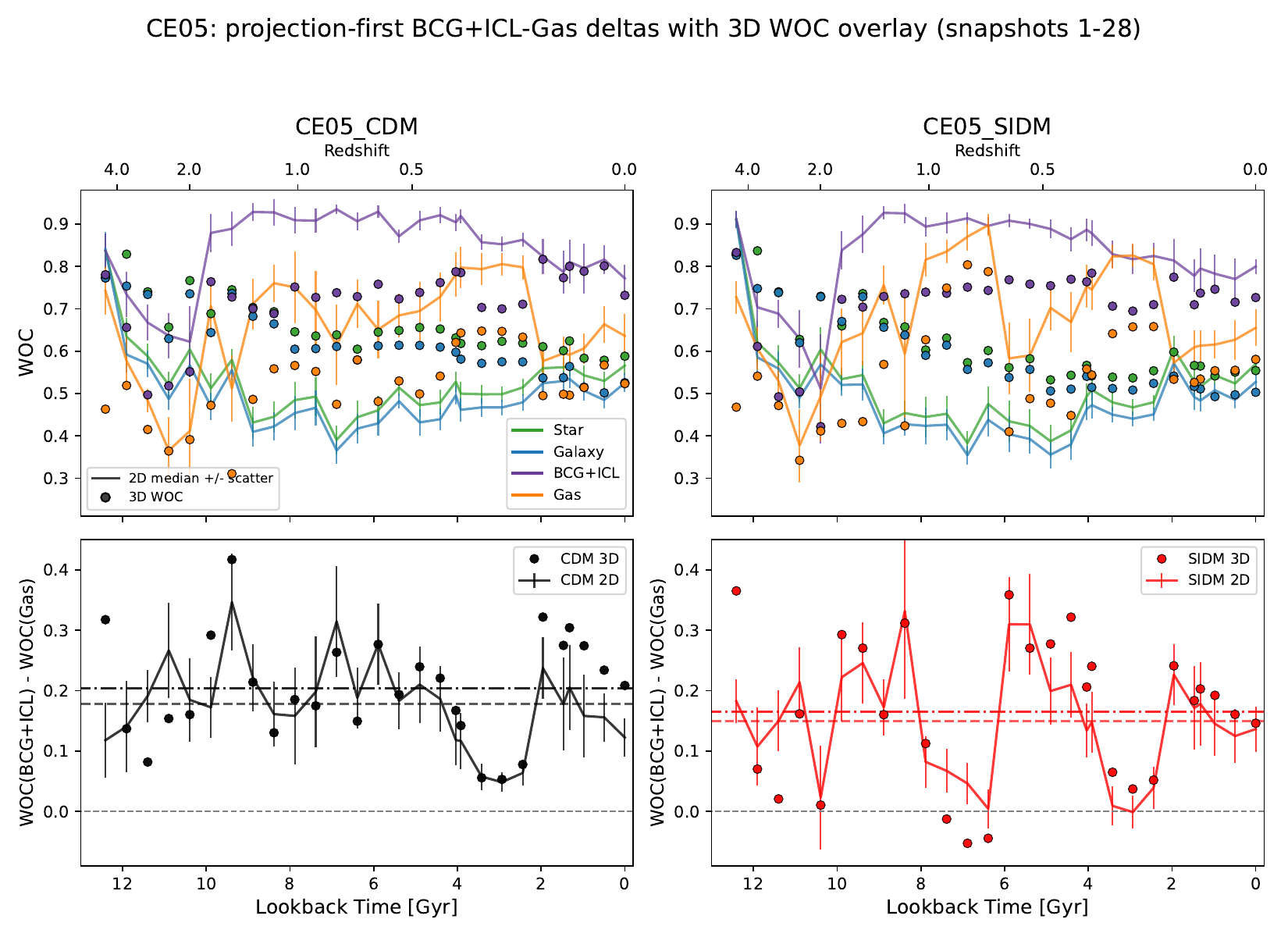}
\caption{ Same as Figure \ref{fig:3Dwoc_ce12} but for cluster \texttt{CE05}.
\label{fig:3Dwoc_ce5}
}

\end{center}
\end{figure}

 In this section we compare the 3D WOC results to those of the 2D analysis in Section~\ref{sec:result}. To ensure a consistent comparison between the 2D and 3D WOC measurements, we adopt the same pixel/voxel scale and smoothing kernel size in both analyses.
In Figures \ref{fig:3Dwoc_ce12} and \ref{fig:3Dwoc_ce5} we can see that for BCG+ICL and gas components, the WOC compared to dark matter is higher in 2D compared to 3D. This can be understood for a 2D projection, where a line-of-sight mismatch could be partially or fully hidden. However, in 3D the mismatch is fully represented. 

However, when we consider all stars and galaxy components, the 3D WOC is rather higher than 2D. This is not necessarily contradictory. It is a sign that projection has two competing effects. WOC is basically comparing high-density dark matter contour regions to high-density component contour regions with matched area/volume. Thus, the question is not only “does projection hide depth mismatch?” but also “does projection preserve the rank ordering and compactness of the component peaks?”

Gas and ICL are extended, diffuse, and can be displaced in 3D from the dark matter structure.
In 2D, material at different depths can overlap in projection. Thus, projection artificially boosts overlap, making 2D WOC higher than 3D.

For all stars and galaxies, the opposite can happen because these components are much more compact and halo-bound.  In 3D, those stellar peaks are often genuinely embedded in high-dark matter voxels: BCG in the central halo, satellites inside subhalos.
In a 2D projection, many compact stellar peaks from satellites get projected into a crowded surface-density field. The stellar contours can become a set of small bright projected knots, while the projected dark matter contour is smoother and broadened by line-of-sight material. That can reduce the matched-contour overlap in 2D even though projection helps diffuse components. Thus, a compact collisionless component can have higher 3D WOC, while a diffuse/misaligned component can have higher 2D WOC.

Although there are differences between the 2- and 3D WOC results, when we come to comparing WOC results of various components the difference is small and negligible, as can be seen in the lower panels of Figures \ref{fig:3Dwoc_ce12} and \ref{fig:3Dwoc_ce5}. We may safely conclude that our conclusions in Section~\ref{sec:analysis} are unaffected by the projected 2D analysis.

\end{document}